\documentclass[aps,showpacs,nofootinbib]{revtex4}
\usepackage{epsf,latexsym}
\usepackage{amsfonts}
\begin{document}

\title{Effective null Raychaudhuri equation}
\author{Alessandro Pesci\footnotetext{e-mail: pesci@bo.infn.it}}
\affiliation
{INFN Bologna, Via Irnerio 46, I-40126 Bologna, Italy}

%new0
\begin{abstract}
The effects on Raychaudhuri's equation
of an intrinsically-discrete
or particle nature of spacetime are investigated.
This is done through the consideration of null congruences
emerging from, or converging to, a generic point of spacetime,
i.e. in geometric circumstances somehow prototypical
of singularity issues.
We do this
from an effective point of view,
that is through a (continuous) description of spacetime modified to embody
the existence of an intrinsic discreteness on the small scale,
this adding to previous results for non-null congruences.
  
Various expressions for the effective
rate of change of expansion
are derived.
They in particular provide finite values
for the limiting effective expansion
and its rate of variation
when approaching the focal point.
Further, this results
in a non-vanishing of
the limiting cross-sectional area itself of the congruence.
\end{abstract}
%\new0

%\pacs{}

\maketitle

$ $
%%%%%%%%%%%%%%%%%%%%%%%%%%%%%%%%%%%%%%%%%%%%%%%%%%%%%%%%%%%%%

Recently,
an effective metric,
or qmetric, bitensor $q_{ab}$ has been introduced \cite{KotE, KotF, StaA},
capable of implementing the existence of an intrisic
discreteness or particle nature of spacetime at the microscopic scale,
while keeping the benefits
of a continuous description for calculus \cite{Pad02}.
$q_{ab}$ acts like a metric
in that it provides a (modified) squared distance between two generic
spacelike or timelike separated events $P$ and $p$ (considered as
base and field point, respectively),
which approaches the squared distance as
%new
of an ordinary
%new
$g_{ab}$ metric
when $P$ and $p$ are far away.
Contrary to a metric however,
the squared distance approaches $\epsilon L^2$
(with $\epsilon = 1 (-1)$
for spacelike (timelike) separation)
in the coincidence limit 
$p \to P$, with $L$ being an invariant length characterizing
the qmetric.

In \cite{PesL} an extension of this qmetric approach
to include the case of null separated events has been considered,
and an expression of $q_{ab}$ for them has been provided.
This case could be directly relevant for the study of horizons.
In the case of null geodesics near a focal point,
this might be exploited for example
to study event horizons at their birth
(described e.g.
in \cite{HawB} (in particular Figure 57),
\cite{MisE} Figure 34.7,
and \cite{ThoB} Box 12.1).  
When these geodesics are meant as histories of ultrarelativistic
or massless particles,
%new
we are led to
%new
singularity
formation issues.
In view of this,
%new
the aim
%new
of this note is to investigate
how the null Raychaudhuri equation
gets modified by intrinsic discreteness
of spacetime, as captured by the qmetric,
near a focal point.

%new1
A wide range of results have been obtained in the past
concerning the study of quantum effects on the Raychaudhuri equation.
We would mention in particular the 
results obtained in Loop Quantum
Gravity/Cosmology (LQG/LQC) \cite{AshA, BojB},
which provide a detailed account,
under isotropic conditions, of the resolution of Schwarzschild's
singularity as well as of the avoidance
of the Big Bang singularity formation.
In a different vein,
the studies originated in \cite{DasA}
are somehow prototypical of attempts to include
quantum effects in Raychaudhuri equation with no reference
to any specific quantum theory of gravity.
These latter studies are successful, as well, in showing that quantum effects
protect against singularity formation.
The present attempt has also no reference to any
definite quantum theory of gravity.
The difference with \cite{DasA} is in the way quantum effects
are introduced:
there, through consideration of quantum trajectories
as in Bohm's pilot wave formulation of quantum mechanics;
here, upon assuming the existence of a finite lower-limit
invariant length $L$ between space- or time-separated events.
The present study elaborates on previous results
concerning the effects $L$ induces on the rate of change
of expansion for timelike/spacelike congruences \cite{KotG}.
%new1

%--------------------------------------------------------------------------

In \cite{KotE, KotF, StaA},
the qmetric is introduced as something which leads to replace the quadratic
distance $\sigma^2(p, P)$ between spacelike/timelike separated events
by an effective distance $[\sigma^2]_q = S_L(\sigma^2)$
dependent on the characterizing scale $L$.
%new
This effective distance is
%new
subject to the requirements
$S_L \to \epsilon L^2$ when $\sigma^2 \to 0$ and
$S_L \sim \sigma^2$ when $\sigma^2/L^2$ is large,
as well as to an additional request in the form of the
effective kernel $[G]_q$ of the d'Alembertian,
namely that $[G]_q(\sigma^2) = G(S_L)$ in all maximally
symmetric spacetimes. This fixes the expression of $q_{ab}(p, P)$
to the form

\begin{eqnarray}\label{qmetric_K}
  q_{ab} = A g_{ab} + \epsilon \ \Big(\frac{1}{\alpha} - A\Big) t_a t_b, 
\end{eqnarray}
where $t^a$ is the normalized tangent vector
($g_{ab} t^a t^b = \epsilon$; $t_a = g_{ab} t^b$) at $p$
to the geodesics connecting $P$ and $p$,
$g_{ab}$ is considered at $p$,
and $\alpha$ and $A$ are functions of $\sigma^2$,
given by

\begin{eqnarray}\label{alpha}
  \alpha = \frac{S_L}{\sigma^2 \ {S'_L}^2},
\end{eqnarray}  

\begin{eqnarray}\label{A}
  A = \frac{S_L}{\sigma^2} \
      \Big(\frac{\Delta}{\Delta_S}\Big)^\frac{2}{D-1}.
\end{eqnarray}
Here the prime symbol indicates differentiation with respect
to $\sigma^2$,
and $\Delta$ is van Vleck determinant
(\cite{vVl, Mor, DeWA, DeWB}; see \cite{Xen, VisA, PPV})

\begin{eqnarray}
  \nonumber
  \Delta(p, P) = - \frac{1}{\sqrt{g(p) g(P)}}
        \det \Big[- \nabla_a^{(p)} \nabla_b^{(P)} \frac{1}{2} \sigma^2(p, P)\Big]
\end{eqnarray}
($g = \det g_{ab}$), and $\Delta_S(p, P) = \Delta({\tilde p}, P)$
with ${\tilde p}$ being that point on the geodesic through $P$ and $p$
(on the same side of $p$) with $\sigma^2({\tilde p}, P) = S_L(p, P)$.

The extension of this approach to include the null case \cite{PesL}
is done shifting the focus of attention from quadratic distance,
which is identically vanishing in this case,
to affine parameterization.
Exploiting the fact that an affine parameter $\lambda$,
assigned with a null geodesics $\gamma$,
is a distance as measured along $\gamma$ by suitable canonical
observers parallelly-transported along it,
the qmetric
is introduced as something
which leads to replace $\lambda(p, P)$ (having $\lambda(P, P) = 0$)
with an effective parameterization $[\lambda]_q = \tilde\lambda(\lambda)$,
which depends on the characterizing scale $L$
(we omit the explicit indication of this dependence).
%new
The effective parameterization has the requirements
%new
$\tilde\lambda \to L$ when $\lambda \to 0$ and
$\tilde\lambda \sim \lambda$ when $\lambda/L$ is large,
as well as the same additional request on the form of the
effective kernel $[G]_q$ of the d'Alembertian as above, specialized
to points on null geodesics.
%new
This last request consists in
what is derived for points null separated from $P$
from requiring $[G]_q(\sigma^2) = G(S_L)$ in all
maximally-symmetric spacetimes.
%new
This gives,
for $q_{ab}(p, P)$
with $P$ and $p$ null separated, the expression 

\begin{eqnarray}
  \nonumber
  q_{ab} = A_\gamma g_{ab} - \Big(\frac{1}{\alpha_\gamma} - A_\gamma\Big)
  l_{\left(a\right.} m_{\left.b\right)},
\end{eqnarray}
with
$
l^a  = \frac{dx^a}{d\lambda}
$
and $m^a$ null with 
$
g_{ab} m^a l^b = -2
$
considered at $p$ (as well as $g_{ab}$ is),
$l_a = g_{ab} l^b$, $m_a = g_{ab} m^b$,
and $\alpha_\gamma$ and $A_\gamma$ functions of $\lambda$
given by

\begin{eqnarray}\label{alpha_gamma}
  \alpha_\gamma =
  \frac{1}{(d\tilde\lambda/d\lambda)^2},
\end{eqnarray}

\begin{eqnarray}
  \nonumber
  A_\gamma = \frac{\tilde\lambda^2}{\lambda^2}
  \Big(\frac{\Delta}{\Delta_S}\Big)^{\frac{2}{D-2}}
  \Big(\frac{d\tilde\lambda}{d\lambda}\Big)^{-\frac{2}{D-2}}.
\end{eqnarray}  
Here
$
\Delta_S(p, P) = \Delta({\tilde p}, P),
$
where $\tilde p$ is that point on $\gamma$ (on the same side of $p$)
which has $\lambda({\tilde p}, P) = \tilde\lambda$ 
with
$
(\partial^a \sigma^2)_{|{\tilde p}} = \partial^a S_L =
2 {\tilde\lambda} l^a_{|{\tilde p}}.
$

The functions $\alpha_\gamma$ and $A_\gamma$ are defined for points
on the null geodesic from $P$
and then only
on the submanifold $\Gamma$ consisting of the null congruence
of all null geodesics emerging from $P$ (considered as base point).
Crucial in the derivation of these expressions,
is considering the d'Alembertian at points of $\Gamma$
in a form which has no derivations
of the vectors tangent to the congruence
taken along directions
outside $\Gamma$ \cite{PesL}. 
This has been accomplished through the following expression for
the d'Alembertian
(meant as applied to a generic function $f(\sigma^2)$
in a maximally-symmetric spacetime)

\begin{eqnarray}
  \nonumber
  \Box f =
  \nabla_a \nabla^a f =
  \big(4 + 2 \lambda \nabla_i l^i\big) \frac{df}{d\sigma^2}
\end{eqnarray}
($i = 1, ..., D-1$ are indices of components on $\Gamma$),
i.e. in terms of a quantity, 
$
\nabla_i l^i = \theta,
$
the expansion of $\Gamma$,
in which all variations are
in $\Gamma$.
Expressions of $[\nabla_i l^i]_q$ have then been readily obtained
as 

\begin{eqnarray}
  \label{div_1}
  [\nabla_i l^i]_q
  &=&
  \nabla_i\Big(\frac{d\lambda}{d\tilde\lambda} l^i\Big) +
  \frac{1}{2} \frac{d\lambda}{d\tilde\lambda} q^{bc} l^a\nabla_a q_{bc}
  \\
  \label{div_2}
  &=&
  \frac{d\lambda}{d\tilde\lambda} \nabla_i l^i -
  \frac{d\lambda}{d\tilde\lambda} \frac{d}{d\lambda}
  \ln \frac{d\lambda}{d\tilde\lambda} +
  \frac{1}{2} (D-2) \frac{d\lambda}{d\tilde\lambda} \frac{d}{d\lambda}
  \ln A_\gamma,
\end{eqnarray}
where $q^{ab}$ is the inverse of $q_{ab}$.
These expressions provide
the expansion $[\theta]_q$ of the null congruence $\Gamma$
according to the qmetric.
%new
The aim
%new
of this brief report,
is to discuss what the associated effective null Raychaudhuri equation
is and to explore both this and
the effective expansion $[\theta]_q$ at coincidence limit
$p \to P$.
The results we obtain refer to a null congruence emerging
from generic $P$, but can equivalently be read as referring
to a null congruence converging to $P$ upon substitution
$\lambda \to -\lambda$, $\tilde\lambda \to -\tilde\lambda$
and $L \to -L$.

We begin by noting that,
if we use of the expressions for $\alpha_\gamma$ and $A_\gamma$
and introduce the quantity

\begin{eqnarray}\label{A_gamma*}
  A_\gamma^* =
  A_\gamma \Big(\frac{d\tilde\lambda}{d\lambda}\Big)^\frac{2}{D-2} =
  \frac{\tilde\lambda^2}{\lambda^2}
  \Big(\frac{\Delta}{\Delta_S}\Big)^\frac{2}{D-2},
\end{eqnarray}
we can recast equation (\ref{div_2}) as

\begin{eqnarray}\label{theta_1}
  [\theta]_q = \sqrt{\alpha_\gamma}
               \Big[\theta + (D-2) \frac{d}{d\lambda} \ln \sqrt{A_\gamma^*}\Big].
\end{eqnarray}
From this,
considering
the derivative of $\theta$ according to the qmetric

\begin{eqnarray}
  \nonumber
  \Big[\frac{d\theta}{d\lambda}\Big]_q
  &=&
  [l^a \nabla_a \theta]_q \\
  \nonumber
  &=&
  [l^a]_q \ \partial_a [\theta]_q \\
  \nonumber
  &=&
  \frac{d\lambda}{d\tilde\lambda} \ l^a \partial_a [\theta]_q \\
  \nonumber
  &=&
  \frac{d\lambda}{d\tilde\lambda} \ \frac{d}{d\lambda} [\theta]_q \\
  \nonumber
  &=&
  \frac{d}{d\tilde\lambda} [\theta]_q,
\end{eqnarray}
we find

\begin{eqnarray}
  \nonumber
  \Big[\frac{d\theta}{d\lambda}\Big]_q
  &=&
  \alpha_\gamma \frac{d\theta}{d\lambda} +
  \frac{1}{2 \sqrt{\alpha_\gamma}} \ [\theta]_q \
  \frac{d\alpha_\gamma}{d\lambda} +
  (D-2) \ \alpha_\gamma \frac{d^2}{d\lambda^2} \ln \sqrt{A_\gamma^*}
  \\
  &=&
  \label{Ray_1}
  \alpha_\gamma \frac{d\theta}{d\lambda} +
  \frac{1}{2} \Big[\theta + (D-2) \frac{d}{d\lambda} \ln \sqrt{A_\gamma^*}\Big]
  \frac{d\alpha_\gamma}{d\lambda} +
  (D-2) \ \alpha_\gamma \frac{d^2}{d\lambda^2} \ln \sqrt{A_\gamma^*}.
\end{eqnarray}
In the 3rd equality above, use has been made of
$
[l^a]_q = dx^a/d\tilde\lambda = (d\lambda/d\tilde\lambda) l^a.
$

Equation (\ref{Ray_1}) is supposed to be the qmetric
rate of change of the expansion
for the null congruence $\Gamma$.
It exhibits quite a close resemblance to the qmetric rate of change
of expansion found in \cite{KotG} for congruences of unit-tangent
spacelike/timelike integral curves emerging from $P$
(eq. (22) in that paper),
which,
when the congruence is specialized to (spacelike/timelike) geodesics
(which is the context
to which the qmetric
(\ref{qmetric_K}) refers to), reads 

\begin{eqnarray}\label{Ray_K}
  \Big[\frac{d\theta}{d\lambda}\Big]_q =
  \alpha \frac{d\theta}{d\lambda} +
  \frac{1}{2} \Big[\theta + (D-1) \frac{d}{d\lambda} \ln \sqrt{A}\Big]
  \frac{d\alpha}{d\lambda} +
  (D-1) \ \alpha \frac{d^2}{d\lambda^2} \ln \sqrt{A},
\end{eqnarray}
where $\alpha$ and $A$ are given in
equations (\ref{alpha}) and (\ref{A}).
We see that equations (\ref{Ray_1}) and (\ref{Ray_K})
are obtained one from the other through
the replacements
$(D-2), \alpha_\gamma, A_\gamma^*
\leftrightarrow (D-1), \alpha, A$.

Making use of the explicit expressions for $\alpha_\gamma$
and $A_\gamma^*$ (equations (\ref{alpha_gamma}) and (\ref{A_gamma*})),
as well as of the convenient expression

\begin{eqnarray}\label{theta_vV}
  \theta = \frac{D-2}{\lambda} - \frac{d}{d\lambda} \ln \Delta 
\end{eqnarray}  
relating the expansion and the van Vleck determinant in
null congruences (\cite{VisA}; see also \cite{PesL}),
expressions (\ref{theta_1}) and (\ref{Ray_1})
of the expansion and of its rate of change
can be given the form

\begin{eqnarray}\label{theta_2}
  [\theta]_q =
  \frac{D-2}{\tilde\lambda} - \frac{d}{d\tilde\lambda} \ln \Delta_S,
\end{eqnarray}

\begin{eqnarray}\label{Ray_2}
  \Big[\frac{d\theta}{d\lambda}\Big]_q =
  - \frac{D-2}{\tilde\lambda^2} - \frac{d^2}{d\tilde\lambda^2} \ln \Delta_S.
\end{eqnarray}

In these (exact) expressions,
any dependence of $[\theta]_q$ and $[d\theta/d\lambda]_q$
on $\alpha_\gamma$ and $A_\gamma^*$ has been translated
into a dependence on $\tilde\lambda$ and $\Delta_S$.
Comparison with equation (\ref{theta_vV}),
and its derivative

\begin{eqnarray}\label{der_theta_vV}
  \frac{d\theta}{d\lambda} =
  - \frac{D-2}{\lambda^2} - \frac{d^2}{d\lambda^2} \ln \Delta,
\end{eqnarray}
shows that the effective expansion and its effective rate of change
at $p$
with $\lambda = \lambda(p, P)$
turn out to be nothing more than
the expansion and its rate of change
evaluated at point $\tilde p$ on the same null geodesic through $P$ and $p$ 
with $\lambda({\tilde p}, P) = \tilde\lambda$.
From

\begin{eqnarray}\label{der_theta_vV_2}
  \frac{d\theta}{d\lambda} =
  - \frac{\theta^2}{D-2}
  - \frac{2}{\lambda} \frac{d}{d\lambda} \ln\Delta
  + \frac{1}{D-2} \Big(\frac{d}{d\lambda} \ln\Delta\Big)^2
  - \frac{d^2}{d\lambda^2} \ln\Delta
\end{eqnarray}
(upon using (\ref{theta_vV}) in (\ref{der_theta_vV})),
accordingly we also get

\begin{eqnarray}\label{Ray_2bis}
  \Big[\frac{d\theta}{d\lambda}\Big]_q =
  - \frac{{[\theta]_q}^2}{D-2}
  - \frac{2}{\tilde\lambda} \frac{d}{d\tilde\lambda} \ln\Delta_S
  + \frac{1}{D-2} \Big(\frac{d}{d\tilde\lambda} \ln\Delta_S\Big)^2
  - \frac{d^2}{d{\tilde\lambda}^2} \ln\Delta_S. 
\end{eqnarray}

This fact makes equations (\ref{theta_2}) and
(\ref{Ray_2}), as well as (\ref{Ray_2bis}),
quite useful when evaluating 
$[\theta]_q$ and $[d\theta/d\lambda]_q$ at coincidence limit.
We find

\begin{eqnarray}
  [\theta]_0
  &\equiv&
  \lim_{\lambda \to 0} [\theta]_q
  \nonumber \\
  &=& 
  \frac{D-2}{L} - \frac{d}{dL} \ln \Delta_L
  \nonumber \\
  &=&
  \frac{D-2}{L} - \frac{1}{3} \ L \ (R_{ab} l^a l^b)_{|P} +
  o\big[L \ (R_{ab} l^a l^b)_{|P}\big]
  \nonumber \\
  &=&
  \frac{D-2}{L} \Big[1 - \frac{1}{3 (D-2)} \delta + o(\delta)\Big]
\end{eqnarray}
and

\begin{eqnarray}
  \Big[\frac{d\theta}{d\lambda}\Big]_0
  &\equiv&
  \lim_{\lambda \to 0}  \Big[\frac{d\theta}{d\lambda}\Big]_q
  \nonumber \\
  &=&
  - \frac{D-2}{L^2} - \frac{d^2}{dL^2} \ln \Delta_L
  \nonumber \\
  &=&
  \frac{d}{dL} \lim_{\lambda \to 0} [\theta]_q
  \nonumber \\
  &=&
  - \frac{D-2}{L^2} - \frac{1}{3} (R_{ab} l^a l^b)_{|P}
  + o\big[(R_{ab} l^a l^b)_{|P}\big]
  \nonumber \\
  &=&
  - \frac{D-2}{L^2} \Big[1 + \frac{1}{3 (D-2)} \delta + o(\delta)\Big],
\end{eqnarray}
as well as

\begin{eqnarray}
  \Big[\frac{d\theta}{d\lambda}\Big]_0 =
    - \frac{{[\theta]_0}^2}{D-2}
  - \frac{2}{L} \frac{d}{dL} \ln\Delta_L
  + \frac{1}{D-2} \Big(\frac{d}{dL} \ln\Delta_L\Big)^2
  - \frac{d^2}{dL^2} \ln\Delta_L,
\end{eqnarray}
where
$\Delta_L$
is defined as
$\Delta_L = \Delta(\bar p, P)$ with $\bar p$ on $\gamma$ such that
$\lambda(\bar p, P) = L$,
and we used of the expansion (\cite{DeWA} and \cite{Xen, VisA, PPV}) 
\begin{eqnarray}
  \nonumber
  \Delta(p, P) = 1 +\frac{1}{6} \lambda^2 (R_{ab}l^a l^b)_{|P}
  + o\Big[\lambda^2 (R_{ab}l^a l^b)_{|P}\Big]
\end{eqnarray}  
of the van Vleck determinant
and put
$\delta \equiv L^2 \ (R_{ab} l^a l^b)_{|P}$
with the expansions useful
when
$\delta \ll 1$;
this sets a maximum allowed value for $(R_{ab} l^a l^b)_{|P}$.
We see that,
whereas classically, i.e. according to $g_{ab}$,
both $\theta$  and $d\theta/d\lambda$ diverge
when $p \to P$ 
(being
$
\theta \sim \frac{D-2}{\lambda}
$
and
$
\frac{d\theta}{d\lambda} \sim -\frac{D-2}{\lambda^2}
$
for $\lambda \to 0$),
according to the qmetric they both remain finite,
the limiting values of $[\theta]_q$ and $[d\theta/d\lambda]_q$
turning out to be the expressions for $\theta$ and $d\theta/d\lambda$
computed at $\lambda = L$.

This adds, and corresponds, to the non-vanishing of the
effective cross-sectional $(D-2)$-dimensional area of $\Gamma$
in the coincidence limit $p \to P$.
Indeed,
from

\begin{eqnarray}\label{effective_volume}
  \big[d^{D-1} V\big]_q
  &=&
  \Big(\frac{\tilde\lambda}{\lambda}\Big)^{D-2} \frac{\Delta}{\Delta_S}
  \ d^{D-2}{\cal A} \ d\lambda
  \nonumber \\
  &\equiv&
  [d^{D-2}{\cal A}]_q \ d\lambda 
\end{eqnarray}
(\cite{PesL}, equation (32), upon using
the explicit expression for $A_\gamma$),
where
$\big[d^{D-1} V\big]_q$ is the effective volume element
and $[d^{D-2}{\cal A}]_q$ the effective cross-sectional area
of the volume element
$d^{D-1} V = d^{D-2}{\cal A} \ d\lambda$ of $\Gamma$,
we get
\begin{eqnarray}\label{cross}
  [d^{D-2}{\cal A}]_0
  &\equiv&
  \lim_{\lambda \to 0} [d^{D-2}{\cal A}]_q
  \nonumber \\
  &=&
  L^{D-2} \frac{1}{\Delta_L} (d\chi)^{D-2},
\end{eqnarray}
where we consider  
as the cross-sectional area element
a $(D-2)$-cube of edge $\lambda d\chi$.
This completes what we were searching for.

If we start now from the classical Raychaudhuri equation as applied
to our (affinely-parameterized) null congruence $\Gamma$, written as

\begin{eqnarray}
  \nonumber
  \frac{d\theta}{d\lambda} = - \frac{1}{D-2} \ \theta^2
  - \sigma_{ab} \ \sigma^{ab} - R_{ab} \ l^a l^b
\end{eqnarray}  
($\sigma_{ab}$ is shear; the twist is vanishing due
to surface-orthogonality),
and use of (\ref{der_theta_vV_2}),
we get

\begin{eqnarray}
  \sigma_{ab} \ \sigma^{ab} + R_{ab} \ l^a l^b =
  \frac{d^2}{d\lambda^2} \ln \Delta +
  \frac{2}{\lambda} \ \frac{d}{d\lambda} \ln \Delta -
  \frac{1}{D-2} \Big(\frac{d}{d\lambda} \ln \Delta\Big)^2,
\end{eqnarray}
and, from (\ref{Ray_2bis}),

\begin{eqnarray}
  [\sigma_{ab} \sigma^{ab}]_q + [R_{ab} l^a l^b]_q =
  \frac{d^2}{d{\tilde\lambda}^2} \ln \Delta_S +
  \frac{2}{\tilde\lambda} \ \frac{d}{d{\tilde\lambda}} \ln \Delta_S -
  \frac{1}{D-2} \Big(\frac{d}{d{\tilde\lambda}} \ln \Delta_S\Big)^2,
\end{eqnarray}
with its coincidence limit

\begin{eqnarray}
  \lim_{\lambda \to 0} \Big([\sigma_{ab} \sigma^{ab}]_q + [R_{ab} l^a l^b]_q\Big) =
  \frac{d^2}{dL^2} \ln \Delta_L +
  \frac{2}{L} \ \frac{d}{dL} \ln \Delta_L -
  \frac{1}{D-2} \Big(\frac{d}{dL} \ln \Delta_L\Big)^2. 
\end{eqnarray}  
In particular, we can read here the expression for $[R_{ab} l^a l^b]_q$
and its coincidence limit in the shearless case.

To conclude, we briefly comment on a consequence
%new
of the above
%new
regarding singularities.
Let us consider the spacetime associated to a spherical layer
of photons, assumed to be pointlike particles,
undergoing spherically symmetric collapse towards a focal point $P$
(we could consider massive particles as well,
but we choose photons
%new
to adhere
%new
to the results presented above).
In our picture,
we can look at this as
a spherically-symmetric
congruence of null geodesics emerging from $P$
and tracked backwards in time, with the further
crucial assumption that
these geodesics are actual histories of photons,
which are then considered as source
of spacetime curvature.
For these circumstances, the classical description
tells us that a singularity unavoidably develops
(this is a sort of prototypical case of singularity formation
in general relativity).
Indeed,
photons
reach $P$ in a finite variation $\Delta\lambda$
of affine parameter,
with diverging energy densities $\rho = E/{\cal A}$
(energy per unit transverse area).
This means that,
in a finite $\Delta\lambda$, photon histories do cease to exist,
while some components  of the Riemann tensor w.r.t. a
basis parallelly-propagated along the geodesics grow without limit,
i.e. we have incomplete geodesics corresponding
to a parallelly-propagated singularity curvature \cite{HawB}.

According to the qmetric description,
in that same $\Delta\lambda$
photon histories keep staying away from $P$ (since the spatial distance
from the actual location $p$ of the photon and $P$
according to any canonical observer at $P$
remains no lower than $L$),
and energy density reaches a maximum insurmountable value
$[\rho]_0 = \lim_{\lambda \to 0} [\rho]_q = E/[{\cal A}]_0$.
Then photon histories do not cease to exist after $\Delta\lambda$
and, using the density $[\rho]_0$ as source of Einstein's equations,
no components of Riemann in a parallelly-propagated basis
are any longer diverging.
In this sense we can say then that the microstructure of spacetime,
as captured by qmetric, removes a classically-blatant
curvature singularity.

Assuming $L$ is as small as orders of Planck's length,
the density $[\rho]_0$ actually challenges the domain of
validity of Einstein's equations and the notion of spacetime,
as can be envisaged by computing
(equation (\ref{cross})) 

\begin{eqnarray}\label{singul}
  [{\cal A}]_0 = 4 \pi L^2 \frac{1}{\Delta_L},
\end{eqnarray}
where
$\Delta_L = \Delta({\bar p}, P)$ is finite
in spite of being the classical metric singular at $P$
when $\lambda = 0$
($\Delta({\bar p}, P)$ is indeed computed for the metric
configuration associated with $\lambda({\bar p}, P) = L$,
that is, clearly, with $\lambda \ne 0$).  
The qmetric thus embodies that, after $\Delta\lambda$,
%new
the photons' spacetime,
instead of becoming singular,
changes its nature
%new
from continuous to discrete
and calls for new
equations,
different from Einstein's,
to rule its evolution.

%new2
At variance with \cite{DasA},
our derivation does not assume a fixed background spacetime.
Indeed, all quantum spacetime effects at $P$ are thought to be
subsumed by the qmetric,
and the photons which go along the null congruence
actively contribute in determining the qmetric at $P$.
Due to the complete generality of our model,
not much can be said about the specific physical mechanisms 
which lead to a finite expansion and a finite rate of change
of it in the coincidence limit.
What we do can say is that this is an effect
of quantum geometry, since this is what the qmetric embodies.
Our point of view is that
in the approach presented here
the specific physical mechanisms in action
could be handled only when we have some hint on how
to modify Einstein's equations
when we are approaching the scale $L$.
For a detailed account of the manner in which
the formation of a singularity is avoided
one should take into account
in an essential manner the influence of the imploding matter
itself on the geometry, and this requires the new field equations.
What this study seems able to say
is simply that at circumstances
in which general relativity requires a singularity formation,
the granular structure of spacetime as captured by the
qmetric requires that no singularity is formed;
this on general grounds,
whatever
the new field equations will be.
%new2

%new3
Upon comparing of these results with those 
found in LQG and LQC \cite{AshA, BojB},
in our opinion it is fair to say that,
if under isotropic symmetry conditions,
the latter are far more
definite and accurate in their predictions
(e.g., in cosmology, the effective Raychaudhuri equation
when followed backwards towards the
initial singularity predicts a vanishing of the expansion
with a change of sign of the rate of change of the latter,
namely a bounce),
i.e. the approach presented here seems to have some
predictive disadvantages or some loss of accuracy.
This however is part of the game. LQG is indeed a specific theory
of quantum gravity. Here instead we remain as generic as possible
when introducing quantum effects on geometry.
Another way to look at this is to consider that
in LQG the quantization of length is an induced concept.
It is a consequence of a quantization procedure
based on general relativity (discretization of the classical theory
and search of a quantum theory corresponding to this discretization)
\cite{RovQ}.
Here, instead, length quantization is meant as
a simple primary concept, a basic
unavoidable (meaning, it should be present in any quantum theory
of gravity)
quantum effect,
and all the discussion is built on this.
When considering singularity issues, the genericity of our approach
has its own point of merit.
Indeed, the results we obtain,
in particular the avoidance of singularity formation,
happen to be absolutely general (and a similar comment could be done
for \cite{DasA}).
From this,
singularity avoidance in quantum gravity
turns out to be on an even firmer ground
in that it
cannot be considered as specific to the quantum theory
of gravity one is considering,
but happens whichever this theory might be.
%new3

{\it Acknowledgements.} I thank S. Chakraborty, D. Kothawala
and G. Venturi
for having read and made suggestions to earlier versions
of the paper.

%%%%%%%%%%%%%%%%%%%%%%%%%%%%%%%%%%%%%%%%%%%%%%%%%%

\end{document}